\documentstyle[12pt,preprint,tighten,aps,epsf]{revtex}
\def\thalf{{\textstyle{\frac{1}{2}}}}

\def\oneth{{\textstyle{\frac{1}{3}}}}
\def\twoth{{\textstyle{\frac{2}{3}}}}
\newcommand{\vm}[1]{\mbox{\bf#1}}
\newcommand{\vmg}[1]{\mbox{\boldmath$#1$}}

\draft

\newcommand{\be}{\begin{eqnarray}}
\newcommand{\ee}{\end{eqnarray}}

\preprint{\normalsize NUC-MINN-01/5-T}

\title{Properties of $\rho$ and $\omega$ Mesons at Finite Temperature and 
Density as Inferred from Experiment}
\author{{\bf V. L. Eletsky$^{1,2}$, M. Belkacem$^{1,3}$, P. J. Ellis$^{1}$
and J. I. Kapusta$^{1}$} \vspace*{0.1in} \\
{\it $^{1}$School of Physics and Astronomy,
University of Minnesota}\\
{\it Minneapolis, MN 55455, USA} \\
{\it $^{2}$Institute for Theoretical and Experimental Physics} \\
{\it B. Cheremushkinskaya 25, Moscow 117218, Russia} \\
{\it $^{3}$Laboratoire de Physique Quantique, Universit\'e Paul Sabatier}\\
{\it 31062 Toulouse Cedex, France} }
\date{\today}

\begin{document}

\maketitle
\thispagestyle{empty}

\begin{abstract}
The mass shift, width broadening, and spectral density for 
the $\rho$ and $\omega$ mesons in a heat bath of nucleons and pions
are calculated using a general formula which relates the self-energy
to the real and imaginary parts of the forward scattering amplitude.
We use experimental data to saturate the scattering amplitude at low 
energies with resonances and include a background 
Pomeron term, while at high energies a Regge parameterization is used.
The real part obtained directly is compared with the result of a dispersion
integral over the imaginary part.
The peaks of the spectral densities are little shifted from 
their vacuum positions, but the widths are considerably increased due to 
collisional broadening. Where possible we compare with the UrQMD model 
and find quite good agreement. At normal nuclear matter density and a
temperature of 150 MeV the spectral density of the $\rho$ meson has a 
width of 345 MeV, while that for the $\omega$ is in the range 90--150 MeV.

\end{abstract}

\pacs{11.10.Wx, 25.75.-q, 12.40.Vv}

\setcounter{page}{1}

\section{Introduction}

The modification of the free space properties of a vector meson
in hadronic or nuclear matter is an important problem which has
attracted much attention. Among the properties of immediate 
interest are the mass shift and width broadening of the particle in a 
medium. Many authors have studied these 
questions for the $\rho$ meson, and also in some cases the $\omega$ meson, at 
zero temperature in equilibrium nuclear matter, see the reviews of Ref.
\cite{rev} and Refs. \cite{ei,weise,kon,post,fri}. More closely related to 
this paper is
the finite temperature work of Rapp {\it et al.} \cite{rapp,rap} who have 
considered the medium modification of the pions comprising the meson, as 
well as additional medium scattering contributions. There have also been 
studies \cite{hag,sch} of $\omega$ and $\rho$ mesons in a pion heat bath,
although we shall see that nucleons produce a larger effect. QCD sum rules 
have also been employed \cite{rev,weise}, but these are tailored to the small 
distance behavior whereas, as Eletsky and Ioffe \cite{ei} have pointed out, 
the self-energy is determined by meson-nucleon scattering at relatively 
large distances of order 1 fm; see also Ref. \cite{mal}.

Many of these works have relied on effective Lagrangians; however,
we would like to adopt as model-independent an approach as possible. 
Therefore we use experimental data to construct the amplitude for 
vector mesons scattering from pions and nucleons.
The low energy region is described in terms of resonances plus background,
while at high energies a Regge model is employed.
In principle the amplitude should be completely determined by the 
data. In practice there are uncertainties because
the data are often inaccurate and incomplete, particularly for the 
$\omega$ meson. It is therefore important to check that the real and 
imaginary parts of our amplitudes approximately satisfy the dispersion 
relation which follows from the analytic properties of the amplitude.
Using our amplitude the in-medium self-energy of the 
$\rho$ and $\omega$ mesons can be calculated at finite temperature and 
density. We use the leading term of the exact self-energy expansion 
\cite{je} which requires that the densities be sufficiently small that 
only single scatterings are important. Where possible we will compare with 
results from the ultra-relativistic quantum molecular dynamics (UrQMD)
model \cite{bel} which has been extensively tested. For the $\rho$ meson this 
paper represents an updated and improved version of earlier work \cite{ele}
hereinafter referred to as EIK.

This paper is organized as follows. In Sec. II we discuss the formalism,
parameters and results for the scattering amplitudes. These are employed in 
Sec. III where the self-energies of the $\rho$ and $\omega$ mesons are 
presented. Concluding remarks are given in Sec. IV.

\section{Scattering Amplitudes}
\subsection{Low Energy Amplitude}

We assume that the self-energies of the isovector $\rho$ and isoscalar 
$\omega$ vector mesons are dominated by 
scattering from the pions and nucleons present in a heat bath. Accordingly
we need four scattering amplitudes. We will adopt the two-component duality 
approach due to Harari \cite{har} (see also Collins \cite{col}) which states 
that while ordinary Reggeons are dual to $s$-channel
resonances, the Pomeron is dual to the background upon which the resonances 
are superimposed. Taking for definiteness the case of a 
$\rho$ meson scattering from particle $a$, we write the forward
scattering amplitude in the c.m. system as
\begin{equation}
f^{\rm cm}_{\rho a}(s)=\frac{1}{2q_{\rm cm}}\sum_R W^R_{\rho a}
\frac{\Gamma_{R\rightarrow\rho a}}{M_R-\sqrt{s}-\thalf i\Gamma_R}
-\frac{q_{\rm cm}r^{\rho a}_P}{4\pi s}
\frac{(1+\exp^{-i\pi\alpha_P})}{\sin\pi\alpha_P}s^{\alpha_P}\;. \label{lef}
\end{equation}
Here the first term involves a sum over a series of Breit-Wigner 
resonances of mass $M_R$ and total width $\Gamma_R$, while the second
term is the Pomeron background contribution which is discussed in Sec. II B
below. No background contribution was included in EIK
\cite{ele}. For the Breit-Wigner term we have used the 
non-relativistic form which amounts to setting
$M_R+\sqrt{s}\simeq2M_R$ in the relativistic denominator 
$M_R^2-s-i\Gamma_R M_R$. This has a negligible effect on the results we 
present. In the usual notation $\sqrt{s}$ is the total c.m. energy and
the magnitude of the c.m. momentum is
\begin{equation}
q_{\rm cm}=\thalf\sqrt{[s-(m_{\rho}+m_{a})^2]
[s-(m_{\rho}-m_{a})^2]}/\sqrt{s} \;. 
\end{equation}
The statistical averaging factor for spin and isospin is 
\begin{equation}
W^R_{\rho a}=\frac{(2s_R+1)}{(2s_\rho+1)(2s_a+1)}
\frac{(2t_R+1)}{(2t_\rho+1)(2t_a+1)}\;,
\end{equation}
in an obvious notation. Since we are averaging over all spin directions we 
shall not distinguish longitudinal and transverse polarizations. The isospin
averaging implies that all charge states of particle $a$ are equally 
populated so there is no medium-induced mixing \cite{abhee} between the $\rho$ 
and $\omega$ mesons. In eq. (\ref{lef}) $\Gamma_{R\rightarrow\rho a}$ represents 
the partial width for the resonance decay into the $\rho a$ channel.
If we denote the c.m. momentum at resonance by $q_{\rm cm}^R$, then
for $q_{\rm cm}\geq q_{\rm cm}^R$ we use the value obtained from the total 
width and the branching ratio on resonance. However the threshold behavior 
of the partial width is known and we incorporate this for
$q_{\rm cm}\leq q_{\rm cm}^R$ by replacing
$\Gamma_{R\rightarrow\rho a}$ by $\Gamma_{R\rightarrow\rho a}
(q_{\rm cm}/q_{\rm cm}^R)^{2l+1}$, where $l$ is the relative angular momentum
between the $\rho$ and the $a$. Since the total width is the sum of the 
partial widths this dependence should be incorporated in $\Gamma_R$. We
do this in the case that $a$ is a pion, but when $a$ is a 
nucleon there are many decay channels and it becomes impractical, so 
we simply take $\Gamma_R$ to be a constant.

For the case of $\rho N$ scattering we use the $N^*$ and $\Delta^*$ 
resonances from Manley and Saleski \cite{man} which are listed in Table I.
These provide a better match onto the high energy region than the
fit of Vrana {\it et al.} \cite{lee}. It is also necessary to include
the $\Delta(1232)$ and the $N(1520)$ subthreshold resonances 
since they make a substantial contribution. In order to estimate the widths
we assume that the vector dominance model is valid, even though it is 
better suited to high energies. This allows us to relate the photon 
and $\rho$ widths. Specifically, since both resonances are close to 
the $\rho N$ threshold, we can write for each of them 
$\Gamma_{\rho N}=q_{{\rm cm}}\gamma_{\rho N}$
and $\Gamma_{\gamma N}=k_{{\rm cm}}\gamma_{\gamma N}$,
where $k_{{\rm cm}}$ is $\gamma N$ c.m. momentum.
Then vector dominance gives 
\begin{equation}
\gamma_{\gamma N} = 4\pi \alpha \frac{1}{g^2_{\rho}}\Biggl ( 1 +
\frac{g^2_{\rho}}{g^2_{\omega}} \Biggr ) \gamma_{\rho N} \;,
\end{equation}
where $\alpha$ is the fine structure constant. For the coupling to the 
photon we take $g_\rho^2/4\pi=2.54$ and $g_\rho^2/g_\omega^2=1/8$.
The value of $\gamma_{\gamma N}$ can be deduced from the decay width and 
the photon branching ratio of the resonances \cite{pdg}. 

For the case of $\omega N$ much less information is available,
although better data is expected in the future \cite{clas}.
Therefore we adopt two extreme models with the expectation that reality lies
somewhere between the two. The first we refer to as the two resonance 
model since, in addition to the
subthreshold $N(1520)$, we include the two resonances reported by 
Manley and Saleski \cite{man}. These are the $N(1900)$ [$\Gamma_R=498$ MeV,
branching ratio to $\omega N$ 0.30] and the $N(2190)$ [$\Gamma_R=547$ MeV,
branching ratio 0.49]. It must be stressed that there 
is uncertainty in these assignments; for example, Vrana {\it et al.} \cite{lee}
report $\omega N$ strength only for the $N(2190)$ with a roughly similar 
width and branching ratio. For the second model, motivated by the fact that 
the $\rho$ and $\omega$ differ only in isospin, we use for the $\omega$ the 
same $T=\thalf$ $N^*$ resonances as for the $\rho$ with the same 
partial widths, except that we omit the $N(1720)$ since it decays 75--80\%
in the $\rho N$ channel \cite{pdg}. In the other cases the errors are 
sufficiently large that similar $\rho$ and $\omega$ decays could be 
accomodated. We refer to this as the multi-resonance model.
We also examined the alternative procedure of adopting
the decay widths in the $\omega N$ channel for the resonances found in 
the quark model calculations of Capstick and Roberts \cite{cap}. We found, 
however, that the cross section was too small for satisfactory matching
onto the high energy part.

Turning now to the $\rho\pi$ amplitude, Eq. (\ref{lef}) indicates that
a Breit-Wigner contribution for $s$-waves in the limit 
$q_{\rm cm}\rightarrow0$ is a constant since a factor of $q_{\rm cm}$ 
is included in the partial width. According to Adler's theorem the 
pion scattering amplitude on any hadronic target vanishes when 
$q_{\rm cm}\rightarrow0$ in the limit of massless pions. In the framework 
of an effective Lagrangian this can be achieved if a derivative coupling is
used for the pion field, $\partial_\mu\vmg{\pi}$. We assume that the term
in the Lagrangian responsible for $\rho\pi\rightarrow a_1(1260)$ involves
$\partial_\mu\vmg{\pi}$ multiplied by the $\rho$-meson field strength 
tensor $\vmg{\rho}^{\mu\nu}$ and $\vm{a}_{\nu}$ for the $a_1$ field.  
This gives an additional factor which should be included for an $s$-wave
partial width of
\begin{equation}
\Biggl ( \frac{s - m^2_{\rho} - m^2_{\pi}}{s_0 - m^2_{\rho} -
m^2_{\pi}}\Biggr )^{\!2} \, ,
\end{equation}
for $s\leq s_0$, where $s_0$ is a normalization point. When  $s\geq s_0$ this 
factor is replaced by unity.
Since this is a soft pion effect it is reasonable to cut it off when 
$q_{\rm cm}\sim1-2m_\pi$, hence we take the normalization point to be
$s_0=(m_\rho+2m_\pi)^2$. We believe this is more reasonable than taking the 
resonance mass for $\sqrt{s_0}$ as in EIK \cite{ele}. The 
analogous factor 
is also introduced for the $h_1(1170)$ resonance. The parameters \cite{pdg}
for these and the other meson resonances included in the calculation 
are listed in Table I. For the $\omega\pi$ amplitude only the $b_1(1235)$
is listed as having appreciable strength \cite{pdg}. We take it to decay 
100\% to $\omega\pi$ with a width of 142 MeV and apply the Adler factor
as outlined above.

\subsection{High Energy Amplitude}

The high energy forward scattering amplitude is known \cite{don} 
to be well approximated by the Regge form
\begin{equation}
f^{\rm cm}_{\rho a}(s)=-\frac{q_{\rm cm}}{4\pi s}\sum_i
\frac{1+\exp^{-i\pi\alpha_i}}{\sin\pi\alpha_i}r^{\rho a}_is^{\alpha_i}\;. 
\label{hef}
\end{equation}
We shall consider a Pomeron term $P$ and a Regge term $P'$. 
In order to obtain the intercept $\alpha_i$ and the residue $r_i$ for the 
$i$'th Regge pole trajectory we use the relation between 
the amplitude and the total cross section given by the optical theorem:
$\sigma_{\rho a}=4\pi {\rm Im} f^{\rm cm}_{\rho a}/q_{\rm cm}$.
High energy scattering is dominated by contributions from individual quarks
-- the additive quark model. Therefore it is reasonable to average over 
charge states and take the cross section 
$\sigma_{\rho N}\simeq\sigma_{\pi N}$. Using
the Particle Data Group listing \cite{pdg} this gives intercepts 
$\alpha_P=1.093$ and $\alpha_{P'}=0.642$ with $r_P^{\rho N}=11.88$ and 
$r_{P'}^{\rho N}=28.59$ (the units yield a cross section in mb with 
energies in GeV). We would like to take 
$\sigma_{\rho\pi}\simeq\sigma_{\pi\pi}$, averaged over charge states.
Of course data for the latter are not available, but for Regge exchange in 
the $t$-channel it is appropriate to invoke factorization \cite{fac}
so that the residue
$r_P^{\rho\pi}\simeq r_P^{\pi\pi}\simeq (r_P^{\pi N})^2/r_P^{NN}=7.508$,
using Ref. \cite{pdg}. Similarly $r_{P'}^{\rho\pi}=12.74$. The intercepts
$\alpha_i$ are universal. These parameters yield cross sections which
are roughly 30\% smaller than in EIK \cite{ele} where
the $\gamma N$ and $\gamma\pi$ cross sections were employed along with vector 
dominance. This, together with the background term in Eq. (\ref{lef}), allows 
us to satisfy the dispersion relation (see below) 
significantly more accurately than with the EIK parameterization \cite{ele}. 

Since the different isospin structure of the $\rho$ and the $\omega$
is expected to be insignificant at high energy, we adopt the same
parameterization for the $\omega\pi$ and $\omega N$ scattering amplitudes 
as for the $\rho\pi$ and $\rho N$ amplitudes. The parameters for the Pomeron 
obtained here are also used 
for the background term in Eq. (\ref{lef}). Note that if the Pomeron 
intercept $\alpha_P$ were exactly 1, the Pomeron amplitude would be pure
imaginary.

\subsection{Results}

\begin{figure}[t]
 \setlength{\epsfxsize}{7.5in}
  \centerline{\epsffile{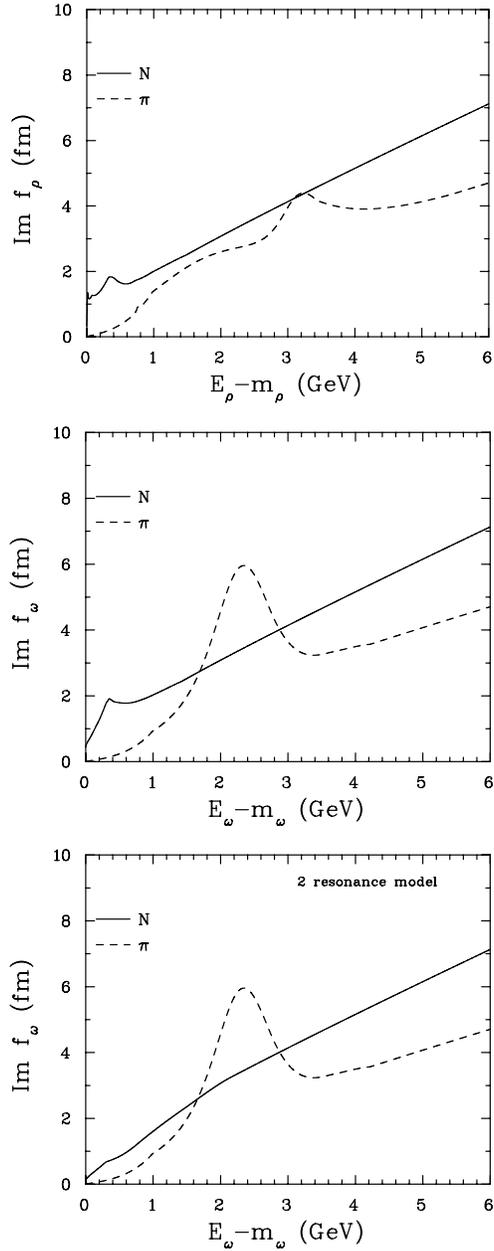}}
\vskip-1.35in
\caption{The imaginary part of the amplitude for $\rho a$ and
$\omega a$ scattering with $a=N,\ \pi$. For the $\omega$ meson we show results 
for the multi-resonance model and the two-resonance model.}
 \label{fig:one}
\end{figure}

\begin{figure}[t]
 \setlength{\epsfxsize}{7.5in}
  \centerline{\epsffile{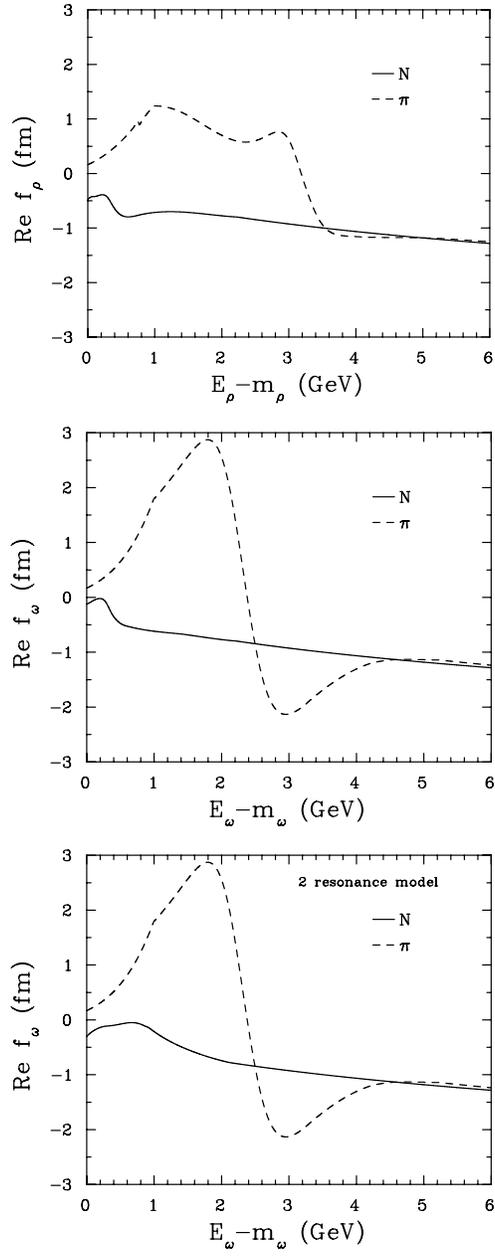}}
\vskip-1.35in
\caption{As for Fig. \ref{fig:one}, but the real parts of the amplitudes.}
 \label{fig:two}
\end{figure}

Since we shall work in the rest frame of the heat bath we will give the 
scattering amplitude for the case that particle $a$ is at rest. This is 
related to the c.m. amplitude by
\begin{equation}
f_{\rho a}(E_\rho)=\frac{\sqrt{s}}{m_a} f^{\rm cm}_{\rho a}(s)\;,
\end{equation}
where 
\begin{equation}
E_\rho-m_\rho=\frac{s-(m_\rho+m_a)^2}{2m_a} \label{kin} \;.
\end{equation}
The imaginary parts of $f_{\rho a}$ and $f_{\omega a}$ are shown in Fig. 
\ref{fig:one}. In most cases the low energy part contains a number of 
overlapping resonances so that the structure is washed out. The exception is 
the case of the $\omega\pi$ amplitude where the single $b_1$ resonance is clearly 
visible (note that this amplitude is the same in the middle and lower panels).
Because of the kinematics, Eq. (\ref{kin}), the resonance region 
ends at $E_\rho-m_\rho\sim1$ GeV for $\rho N$ and $\sim4$ GeV for 
$\rho\pi$ and it is matched onto the Regge part slightly beyond these 
points. At low energies the $\rho N$ amplitude is of similar magnitude 
to the $\omega N$ amplitude in the multi-resonance model, but it is 
much smaller in the two-resonance model. This is less marked for the 
real part of the amplitude, given in Fig. \ref{fig:two}, where the two 
$\omega N$ amplitudes are more similar and both are smaller in magnitude than 
the $\rho N$ amplitude in this resonance region. The pion scattering amplitudes 
display the change in sign expected for Breit-Wigner resonances. This 
is not seen in the nucleon case because of the subthreshold resonances 
included here. These are neglected by Kondratyuk {\it et al.} \cite{kon}
which may be the reason that their $\rho N$ amplitude becomes slightly positive 
at small momenta; it is also somewhat larger in magnitude at large momenta. 
They obtained their result from a dispersion integral over an imaginary 
amplitude constructed from resonances at low energy and vector dominance 
together with photon cross sections at high energy.

\begin{figure}[t]
 \setlength{\epsfxsize}{7.5in}
  \centerline{\epsffile{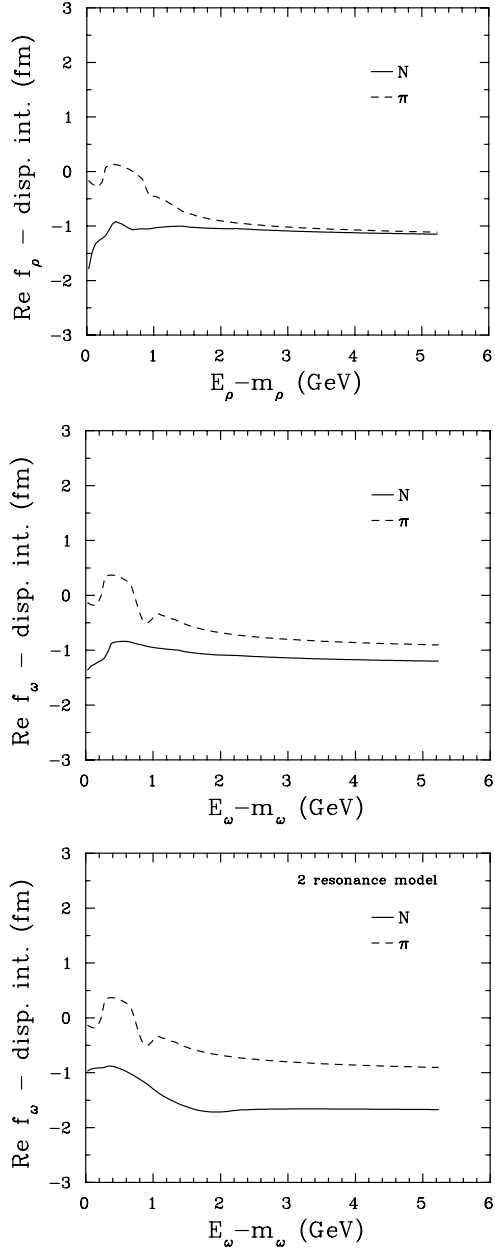}}
\vskip-1.35in
\caption{Difference between the real part of the amplitudes given in Fig. 
\ref{fig:two} and those deduced from the imaginary parts of 
Fig. \ref{fig:one} via the dispersion relation.}
 \label{fig:three}
\end{figure}

The scattering amplitude should obey a once-subtracted dispersion relation
relating the real part to a principal value integral over the imaginary 
part:
\begin{equation}
{\rm Re}f_{\rho a}(E_\rho)={\rm Re}f_{\rho a}(0)
+\frac{2E_\rho^2}{\pi}\:{\rm P.V.}\!\int\limits_{m_\rho}^\infty
\frac{{\rm Im}f_{\rho a}(E')dE'}{E'(E^{\prime2}-E_\rho^2)} \label{di} \;.
\end{equation}
Thus one can compare the analytical real part of Secs. II A and B with the
result from Eq. (\ref{di}); the difference should be the constant 
${\rm Re}f_{\rho a}(0)$. This does not hold if one uses the Regge form for 
$f$ at all energies, the difference only becomes exactly constant if 
the lower limit of the integration is arbitarily taken to be the point 
where $s=0$ \cite{col}.
Alternatively, if one assumes a pure resonance form for the amplitude the 
aforementioned difference is not constant either. In both cases noticeable
deviations from constancy start to appear at energies $E_\rho-m_\rho$ below
about 2 GeV. This trend is also seen for the differences when the actual 
amplitudes are used, as displayed in Fig. \ref{fig:three}. The nucleon 
amplitudes give 
the most reasonable account of the dispersion relation, with the $\omega N$
two-resonance case showing  a larger deviation from constancy than 
the other two cases. For pion scattering the deviations are larger, although 
it should be borne in mind that the amplitudes themselves are larger too.
Of course one would not expect phenomenological approximations to precisely 
obey the stringent constraints which follow from the analytic properties of the 
amplitude, and in that light we view the results in Fig. \ref{fig:three} as 
reasonable. We remark that we have considered variations in the parameters 
involved in the amplitude and have not obtained improvement. In particular,
omission of the background Pomeron term in Eq. (\ref{lef}) gives much worse 
results.

\begin{figure}[t]
 \setlength{\epsfxsize}{7.5in}
  \centerline{\epsffile{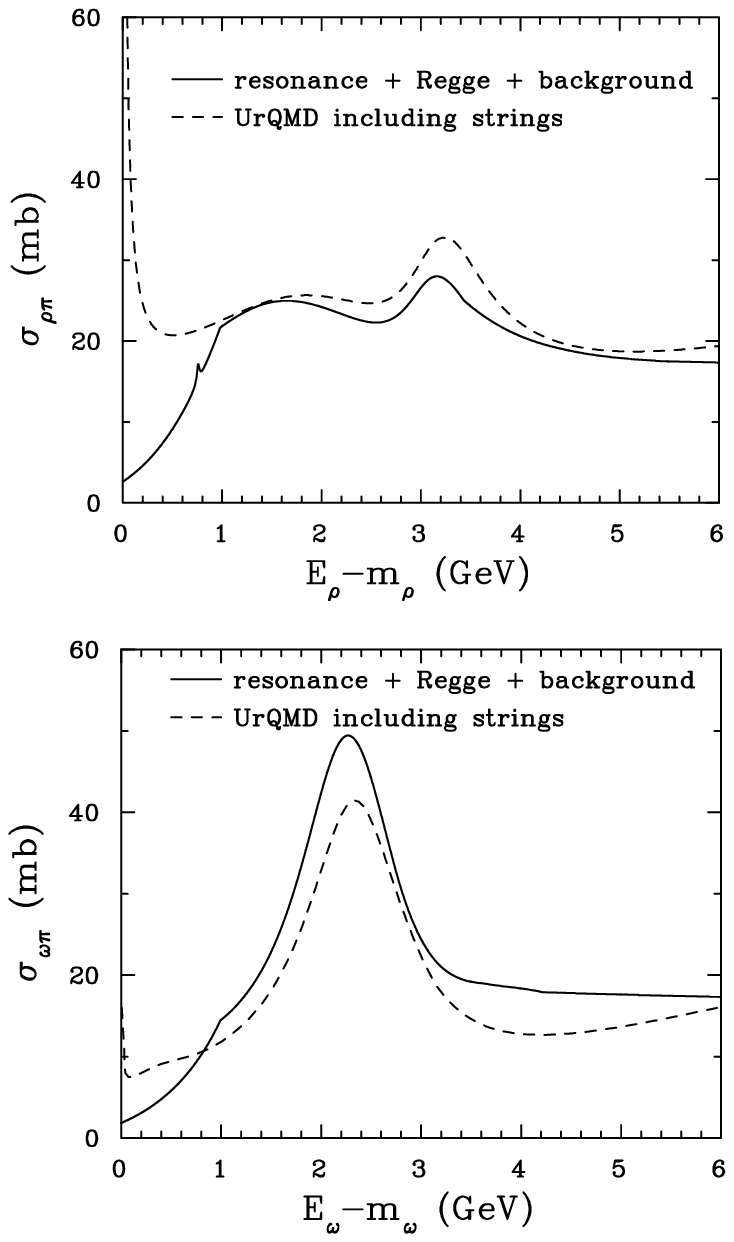}}
\vskip-1.35in
\caption{Cross sections for vector mesons scattering from 
pions: comparison of the present results with those of the UrQMD 
model.}
 \label{fig:four}
\end{figure}

\begin{figure}[t]
 \setlength{\epsfxsize}{7.5in}
  \centerline{\epsffile{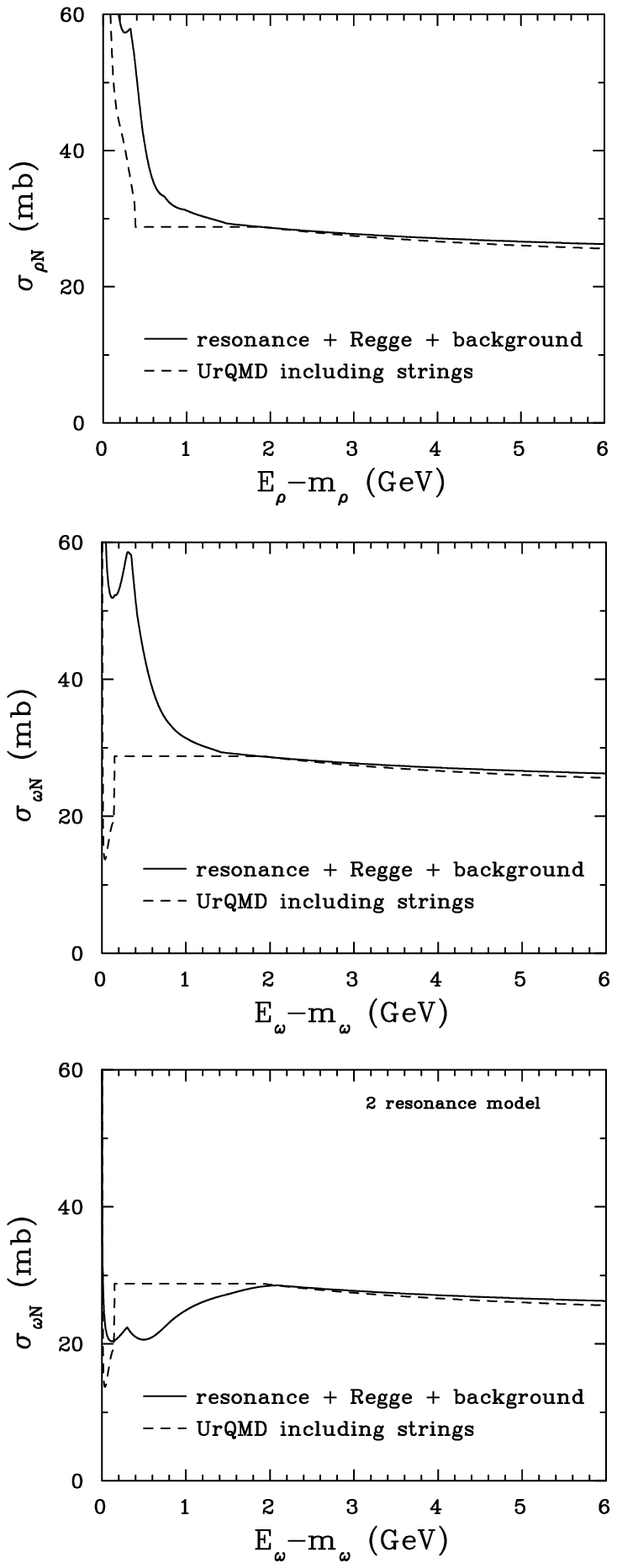}}
\vskip-1.35in
\caption{Cross sections for vector mesons scattering from 
nucleons: comparison of the present results with those of the UrQMD 
model.}
 \label{fig:five}
\end{figure}

There are inevitable uncertainties in a phenomenological parameterization 
of a scattering amplitude so it is useful to compare with other work.
Here we contrast total cross sections calculated from the imaginary parts 
of the amplitudes discussed above with those used in the UrQMD model
\cite{bel}. The latter employs
a resonance description at the lower energies without, however, a background 
term. At the higher energies the CERN--HERA parameterizations \cite{pdg}
are used, together with the additive quark model, leading to color string 
excitations. Comparison 
of the cross sections for scattering from pions in Fig. \ref{fig:four} shows
quite close agreement except at the lowest energies. Here the UrQMD cross 
sections increase because no factor of $(q_{\rm cm}/q_{\rm cm}^R)^{2l+1}$
is included in the width, nor is the Adler factor included. Since 
physically the cross section should go to zero in the chiral limit of 
massless pions we prefer our result where the cross sections are small.
Note that precisely at threshold, $\sqrt{s}=m_\rho+m_\pi$, both approaches 
give a divergent cross section which, however, is of no consequence for 
the calculation of the self energies. The corresponding results for the 
nucleon cross sections are given in Fig. \ref{fig:five}. Again there is very 
close agreement at high energies, but less good agreement at low energies.
For the $\rho$ the basic difference is that UrQMD joins the string region 
to the resonance region at a lower energy. In fact our cross section compares 
better with that of Kondratyuk {\it et al.} \cite{kon}. For the $\omega$ cross 
section only the $N(1900)$ resonance is included in the UrQMD model,
whereas we also include the $N(2190)$. This can be seen rather clearly in 
the lower panel for the two-resonance model. Naturally our multi-resonance 
model for the $\omega$ bears little resemblance to UrQMD (middle panel)
at low energies, being closer to the $\rho N$ case. Apart from this we would 
say that there is broad agreement between UrQMD and the present results.

\section{Self-energies of the Vector Mesons}

For a $\rho$ meson scattering from hadron $a$ in the medium the contribution 
to the retarded self-energy \cite{je,ele} is:
\begin{equation}
\Pi_{\rho a}(E,p) = - 4\pi \int \frac{d^3k}{(2\pi)^3} \,
n_a(\omega) \, \frac{\sqrt{s}}{\omega}
 \, f_{\rho a}^{\rm cm}(s) \;, \label{self}
\end{equation}
where $E$ and $p$ are the energy and momentum of the $\rho$ meson,
$\omega^2 = m_a^2 + k^2$, and 
$n_a$ is either a Bose-Einstein or Fermi-Dirac occupation number as
appropriate for particle $a$. If the self-energy is evaluated on shell in 
the rest frame of $a$ it is possible to do all the angular integrations,
giving
\begin{equation}
\Pi_{\rho a}(p) = - \frac{m_\rho m_a T}{\pi p} \int\limits_{m_a}^{\infty}
d\omega \ln\left[\frac{1-\exp(-\omega_+/T)}{1-\exp(-\omega_-/T)}
\right] f_{\rho a}\left(\frac{m_\rho \omega}{m_a}\right)\;,
\end{equation}
where $\omega_{\pm} = (E \omega \pm pk)/m_\rho$ and $a$ is a boson.
If $a$ is a fermion $\omega_{\pm}$ has an additional chemical potential 
contribution $-\mu$ and the argument of the logarithm becomes
$[1+\exp(-\omega_-/T)]/[1+\exp(-\omega_+/T)]$.

The total self-energy is given by summing over all target species and 
including the vacuum contribution 
\begin{equation}
\Pi_{\rho}^{\rm tot}(E,p) =\Pi_{\rho}^{\rm vac}(M) + 
\Pi_{\rho \pi}(p) + \Pi_{\rho N}(p)\;.
\end{equation}
Here the vacuum part of $\Pi$ can only depend on the invariant mass,
$M = \sqrt{E^2 - p^2}$, whereas the matter parts can in principle depend on
$E$ and $p$ separately.  However, in the approximation we are
using the scattering amplitudes are of necessity evaluated on
the mass shell of the $\rho$ meson.  This means that the matter parts
only depend on $p$ because $M$ is fixed at $m_{\rho}$.  
The dispersion relation is determined from the poles of the propagator
with the self-energy evaluated on shell, {\it i.e.} $M=m_{\rho}$.
Taking again for definiteness the case of the $\rho$ we have
\begin{equation}
E^2 = m_{\rho}^2 + p^2 +\Pi_{\rho}^{\rm tot}(p)\;.
\end{equation}
Since the self-energy has real and imaginary parts so does
$E(p) = E_R(p) -i \Gamma(p)/2$.  
The width is given by 
\begin{equation}
\Gamma(p) = - {\rm Im}\Pi_{\rho}^{\rm tot}(p)/E_R(p) \, ,
\end{equation}
with
\begin{equation}
2E_R^2(p) = p^2+m_{\rho}^2+{\rm Re}\Pi_{\rho}^{\rm tot}(p)
+\sqrt{[p^2+m_{\rho}^2+{\rm Re}\Pi_{\rho}^{\rm tot}(p)]^2  
+[{\rm Im}\Pi_{\rho}^{\rm tot}(p)]^2}   \; . 
\end{equation}
The width of the $\rho$-meson in vacuum, $\Gamma_{\rho}^{\rm vac}
= - {\rm Im}\Pi_{\rho}^{\rm vac}/m_{\rho}$, is 150 MeV (the width of the 
$\omega$-meson in vacuum is 8.4 MeV).
We define the mass shift to be
\begin{equation}
\Delta m_{\rho}(p) = \sqrt{m_{\rho}^2+{\rm Re}\Pi_{\rho}^{\rm tot}(p)} 
- m_{\rho} \;. 
\end{equation}

We assume that the hadronic matter can be considered to be a weakly 
interacting gas of pions and nucleons. In order to 
test this assumption we have run the 
UrQMD code in a box for baryon densities up to twice normal nuclear matter 
density at temperatures up to 150 MeV. The results show that more than 95\% 
of all $\rho$-meson scatterings occur from pions and nucleons so that it is 
reasonable to focus on these interactions. We will consider
nucleon densities of $n_N=$0, 1 and 2 in units of equilibrium nuclear matter 
density ($n_0=0.16$ nucleons/fm$^3$). We assume that the system is in thermal 
equilibrium with a temperature 
below 170 MeV so that hadrons are the approriate degrees of freedom rather 
than quarks and gluons. For densities of 1 and 2
the nucleon chemical potentials are, respectively, 747 and 821 MeV at 
$T$ = 100 MeV, and 543 and 650 MeV at $T$ = 150 MeV.  Anti-nucleons are not 
included.

\begin{figure}[t]
 \setlength{\epsfxsize}{7.5in}
  \centerline{\epsffile{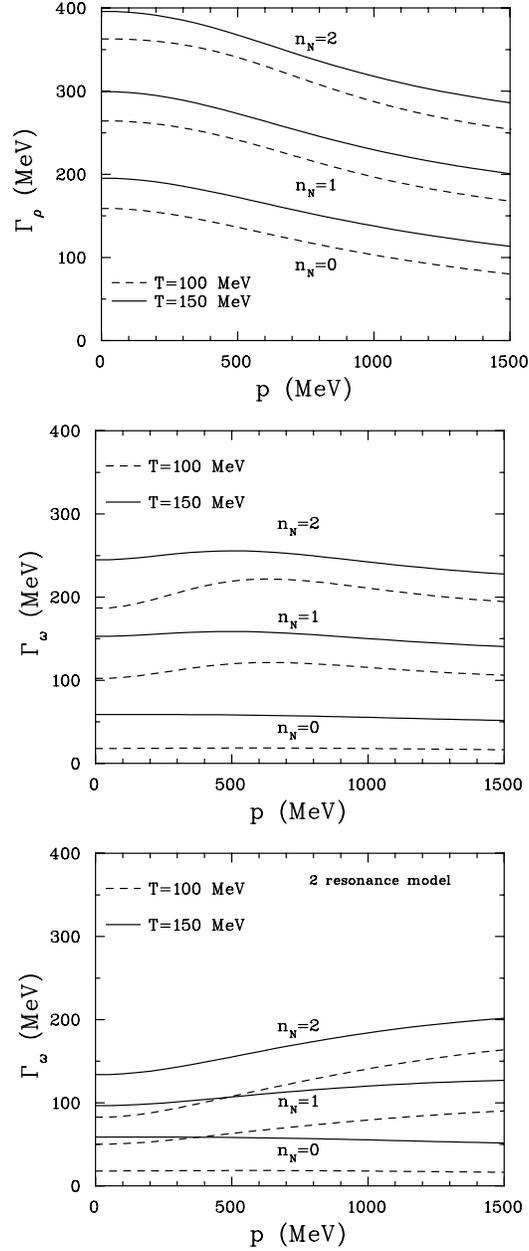}}
\vskip-1.35in
\caption{The vector meson widths as a function of momentum $p$. Results 
are shown for nucleon densities of $0,\ n_0$ and $2n_0$ (where equilibrium 
nuclear matter density $n_0=0.16$ fm$^{-3}$) and temperatures of 100 and 
150 MeV. For the $\omega$ meson results are given for the multi-resonance 
and the two-resonance models.}
 \label{fig:six}
\end{figure}

The vector meson widths are shown as a function of momentum in Fig. 
\ref{fig:six} for two temperatures and three nucleon densities.
Note that the widths given here are defined to be in the rest frame of 
the thermal system. (The present results replace those of Eletsky and 
Kapusta \cite{ele} since the weighting of the pion contribution was too 
small there due to a computer code error. The nucleon contributions still 
dominate, however.) For $\Gamma_\rho$ the $n_N=0$ results are little 
changed from the vacuum value until temperatures of the order of the pion 
mass are reached. At $T=150$ MeV the width generated by collisions with 
pions is about 50 MeV. This is a factor of two larger than obtained by Haglin 
\cite{hag} using an effective Lagrangian, but a little less than the 80 MeV 
reported by Rapp and Gale 
\cite{rapp}. Interactions with nucleons give a 100 MeV contribution to the 
width at $n_N=1$, similar to the zero temperature estimate
of Kondratyuk {\it et al.} \cite{kon}, and about twice that at $n_N=2$.
Thus at the highest temperatures and densities the width is 2--3 times the 
vacuum value and is becoming comparable to the mass. The middle and lower 
panels of Fig. \ref{fig:six} are the same for $n_N=0$ since nucleons are 
not involved in this case. Here the effect of increasing the 
temperature, and therefore the pion 
density, is much more marked than for the $\rho$ since the vacuum width of the 
$\omega$ is so small. At $T=150$ MeV the width is about 50 MeV which is 
similar to the value obtained by Schneider and Weise \cite{sch} in
an effective Lagrangian approach, but a 
factor of two larger than given by Haglin \cite{hag} . When the 
nucleon density is non-zero we expect nature to lie somewhere between the 
larger widths given by the multi-resonance model (middle panel) and the 
smaller widths given by the two-resonance model (lower panel). 
The functional dependence on $p$ differs in the two cases. However, for a 
temperature of 150 MeV and $n_N=1$, $\Gamma_\omega$ is expected to lie between 
100 and 150 MeV. This is an enhancement of the vacuum width by a factor of 
12--18, which is in line with Rapp's estimate \cite {rap} of a factor of 
20 at a slightly higher temperature of 180 MeV.

\begin{figure}[t]
 \setlength{\epsfxsize}{7.5in}
  \centerline{\epsffile{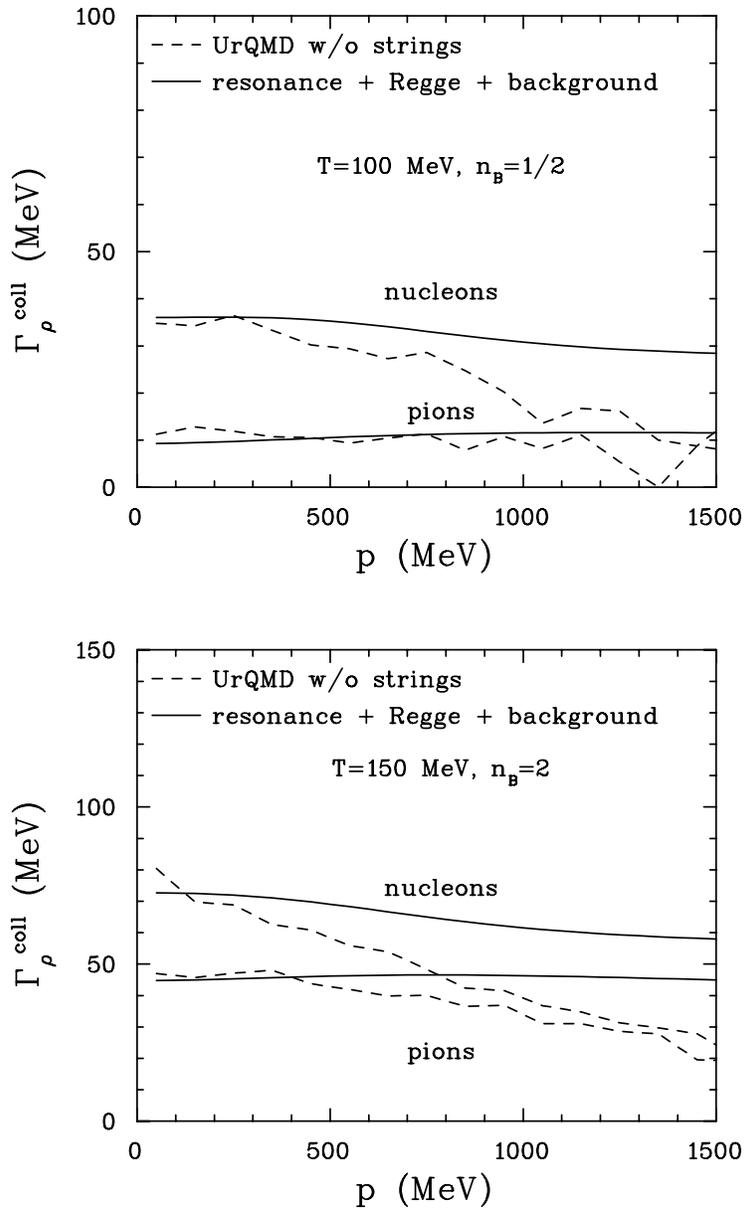}}
\vskip-1.35in
\caption{Comparison of the present results with those of the UrQMD
model (without strings) for the widths generated by collisions with pions
or nucleons. The temperatures and baryon densities for the two cases
are indicated.}
 \label{fig:seven}
\end{figure}

In the UrQMD model collisional widths can be obtained by 
allowing a volume of matter to come to equilibrium at a given temperature 
and baryon density \cite{bel2}. Then the average time between collisions of
a $\rho$ meson with a given species, $N$ or $\pi$,
can be determined. The reciprocal of this gives the width due to 
collisional broadening (in units with $\hbar=1$). In order for the notion of 
thermodynamic equilibrium to make sense detailed balance must hold. 
Therefore, for present purposes, it is necessary to drop the string 
contribution and retain only the resonance contribution \cite{bel2}. Thus the 
results should be most reliable at low momenta. We show the UrQMD 
results for the collisional broadening due to scattering from pions and 
nucleons separately in Fig. \ref{fig:seven}. They are compared with the 
results discussed above for two representative cases of baryon density, $n_B$, 
and temperature. For $n_B=\thalf$ with $T=100$ MeV and $n_B=2$ with $T=150$ MeV, 
the baryon chemical potentials are 630 and 479 MeV, respectively, which 
correspond to nucleon densities $n_N\simeq\oneth$ and $n_N\simeq\twoth$, with 
all densities in units of $n_0$. The present results agree quite nicely with 
UrQMD at low momenta, suggesting that interference between sequential 
scatterings can be ignored at these temperatures and densities as we have done. 
The deviations at larger momenta give some measure of the role played by the 
high energy Regge part of the scattering amplitude.

\begin{figure}[t]
 \setlength{\epsfxsize}{7.5in}
  \centerline{\epsffile{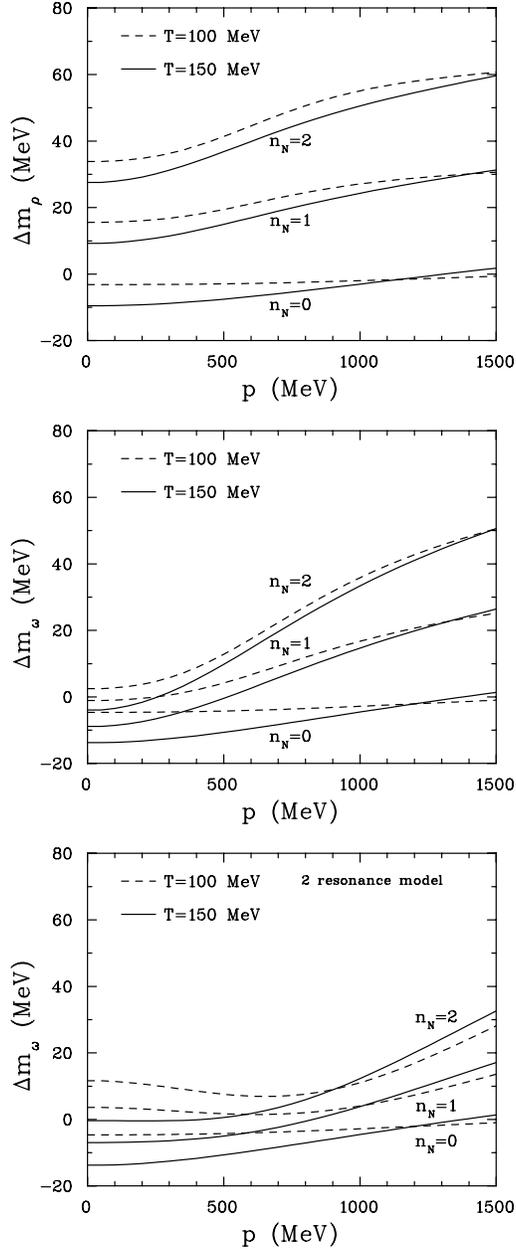}}
\vskip-1.35in
\caption{As for Fig. \ref{fig:six}, but the vector meson mass shifts.}
 \label{fig:eight}
\end{figure}

The mass shifts for the vector mesons are displayed in Fig. \ref{fig:eight}. 
They turn out to be quite small, at most a few tens of MeV. For both the $\rho$ 
and the $\omega$ mesons the interaction with pions alone ($n_N=0$) produces a 
small negative $\Delta m$, while the introduction of nucleons gives a 
positive contribution. For the two $\omega$ models at zero momentum
$\Delta m_\omega$ is in the range $-15$ to $+15$ MeV. On the other hand 
at $p=1500$ MeV for $n_N=2$, $\Delta m_\omega$ is 30 MeV in the two-resonance 
model and 50 MeV in the multi-resonance model, somewhat smaller than 
$\Delta m_\rho=60$ MeV. These trends and numbers for the vector meson mass 
shifts  are roughly consistent with other analyses 
\cite{post,rapp,rap,sch}. However, in nuclear matter at zero temperature 
the coupled-channel calculation of Friman {\it et al.} \cite{fri} gives
larger shifts and, for the $\omega$, so does the chiral approach of Klingl 
{\it et al.} \cite{weise}.

The rate of dilepton production is directly proportional to the
imaginary part of the photon self-energy \cite{mt,w} which is
itself proportional to the imaginary part of the $\rho$ meson
propagator because of vector meson dominance \cite{gs,gk}.
\begin{equation}
E_+ E_- \frac{dR}{d^3p_+ d^3p_-} \propto
\frac{-{\rm Im} \Pi_{\rho}^{\rm tot}}{[M^2 - m_{\rho}^2 
- {\rm Re} \Pi_{\rho}^{\rm tot}]^2 + [{\rm Im}\Pi_{\rho}^{\rm tot}]^2}\;,
\end{equation}
where, as before, $M$ is the invariant mass.
For the $\rho$-meson the vacuum part $\Pi_\rho^{\rm vac}$ can be obtained 
from the Gounaris-Sakurai formula \cite{gs,gk}.  This formula gives a very 
good description of the
pion electromagnetic form factor, as measured in $e^+e^-$ annihilation
\cite{ss}, up to 1 GeV apart from a small mixing with the $\omega$ meson
which we are ignoring in this paper.
\begin{eqnarray}
{\rm Re}\Pi_{\rho}^{\rm vac} &=& \frac{g_{\rho}^2 M^2}{48\pi^2}
\left[ \left(1-4m_{\pi}^2/M^2\right)^{3/2}
\ln\left|\frac{1+\sqrt{1-4m_{\pi}^2/M^2}}{1-\sqrt{1-4m_{\pi}^2/M^2}}
\right| +8m_{\pi}^2\left(\frac{1}{M^2}-\frac{1}{m_{\rho}^2}\right)
\right.  \nonumber \\
&-&\left. 2\left(\frac{p_0}{\omega_0}\right)^3 \ln \left(\frac{\omega_0+p_0}
{m_{\pi}}\right) \right] \;, \\
{\rm Im} \Pi_{\rho}^{\rm vac} &=& -\frac{g_{\rho}^2 M^2}{48\pi}
\left(1-4m_{\pi}^2/M^2\right)^{3/2} \;.
\end{eqnarray}
Here $2\omega_0 = m_{\rho} = 2\sqrt{m_{\pi}^2+p_0^2}$.  The
vacuum width is $\Gamma_{\rho}^{\rm vac} = (g_{\rho}^2/48\pi)
m_{\rho} (p_0/\omega_0)^3$ and the real part vanishes on shell.
Since the vacuum decay of the $\omega$ into three pions is more 
complicated, while the width is tiny, we simply treat it as a constant
except for the application of a non-relativistic phase space factor
$[(M^2-9m_\pi^2)/(m_\omega^2-9m_\pi^2)]^2$ from threshold to $M=m_\omega$.
A possible real vacuum contribution is ignored.

\begin{figure}[t]
 \setlength{\epsfxsize}{7.5in}
  \centerline{\epsffile{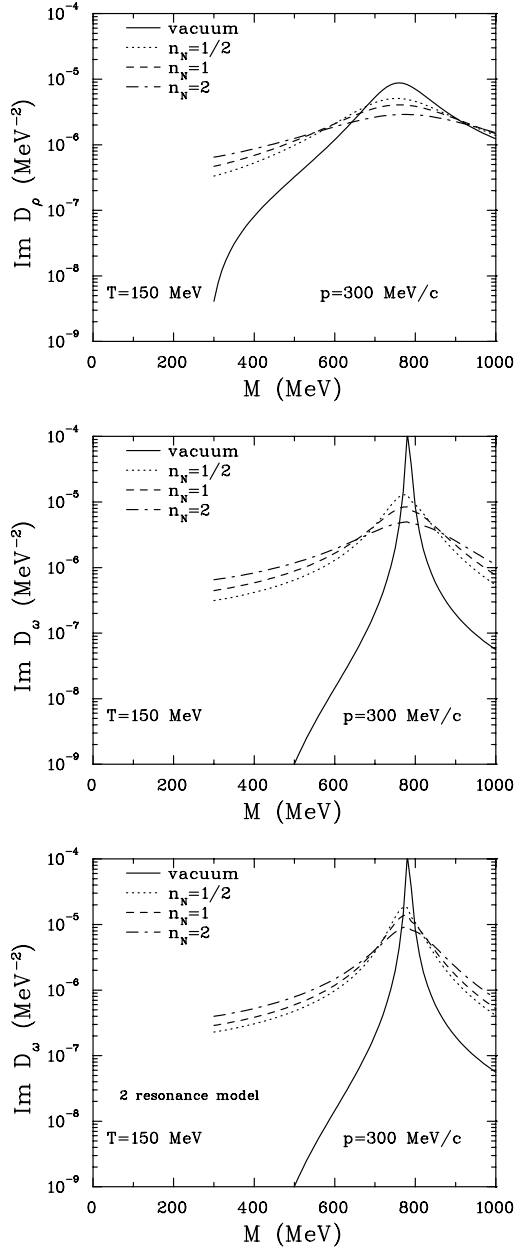}}
\vskip-1.35in
\caption{The imaginary part of the vector meson propagators as a
function of invariant mass for a momentum of 300 MeV/c and a temperature of 
150 MeV. Results are shown for the vacuum and nucleon densities of 
$\thalf n_0,\ n_0$ and $2n_0$. For the $\omega$ meson results are given for 
the multi-resonance and the two-resonance models.}
 \label{fig:nine}
\end{figure}

The imaginary part of the propagator, proportional to the spectral density, 
is plotted as a function of $M$ in Fig. \ref{fig:nine} for
a temperature of 150 MeV. Pions alone have a small effect on the spectral 
density so we display results at $n_N=\thalf,\ 1$ and 2.
These parameters are characteristic of the final stages
of a high energy heavy ion collision.  As seen from Fig. \ref{fig:nine} there 
is little change in the position of the peak, but the spectral density 
is greatly broadened. (In nuclear matter at $T=0$ Refs. \cite{post,fri} 
obtain a more complicated structure.) Figure \ref{fig:nine} shows that for
$n_N=1$ the width of the $\rho$ peak (full width, 
half maximum) is 345 MeV which is becoming comparable to the mass of $\rho$
meson and is consistent with the results of Rapp \cite{rap}. For the $\omega$ 
meson at this density the peak width is 150 MeV in the multi-resonance model 
and 90 MeV for the two-resonance model, while Rapp's width is intermediate 
between these values. 

\section{Conclusions}

In this paper we have described the scattering amplitudes for $\rho$ and 
$\omega$-mesons in terms of resonances plus background at low energies matched
onto a Regge form at high energies (our amplitudes are available upon 
request). The parameters were taken from 
experimental data in order to be as model independent as possible. Of 
course the data are imperfect, particularly for the $\omega$ meson where we 
adopted two extreme models with reality expected to lie somewhere between 
the two. Assuming that only single scatterings are important, as appears to 
be justified by comparison with the UrQMD results, it is then 
straightforward to obtain the self-energy at finite density and 
temperature.

Our results indicate for the shift in the pole mass a negative contribution 
from interactions with pions and a positive contribution from interactions 
with nucleons. The net result is small, at most a few tens of MeV. Thus the 
peak of the spectral density is little shifted, but the width is 
increased considerably due to collisions in the medium. Collisions with 
nucleons dominate, but the effect of pions is not negligible. At equilibrium 
nuclear matter density and a temperature of 150 MeV the width of the 
spectral density is 345 MeV 
for the $\rho$ meson, about twice the vacuum width. For the $\omega$ meson 
the width is expected to lie between the values of 90 MeV and 150 MeV given 
by our two extreme models, a considerable change from the vacuum width of 
just 8.4 MeV. Where possible we have compared with the UrQMD model and 
found quite reasonable agreement. Our results are also quite consistent with 
those of Rapp and coworkers \cite{rapp,rap} and Schneider and Weise 
\cite{sch}.

The next step is to use our results in a space-time model of the 
evolution of matter in high energy heavy ion collisions. This will allow us 
to study the extent to which our spectral densities are able to reproduce
the observed $e^+e^-$ mass spectra \cite{data}. Such work is underway.

\section*{Acknowledgments}

We are indebted to B. L. Ioffe for valuable discussions.
This work was supported in part by the US Department of Energy grant 
DE-FG02-87ER40382 and in part by the US Civilian Research
\& Development Foundation grant RP2-2247.

\begin{table}
\caption{Baryon and Meson Resonances Included in the $\rho$ Amplitude}
\begin{tabular}{|l|ccc|}
Resonance&Mass (GeV)&Width (GeV)&
Branching ratio ($\rho N$ or $\rho\pi$)\\ \hline
$N(1700)$&1.737&0.249&0.13\\
$N(1720)$&1.717&0.383&0.87\\
$N(1900)$&1.879&0.498&0.44\\
$N(2000)$&1.903&0.494&0.60\\
$N(2080)$&1.804&0.447&0.26\\
$N(2090)$&1.928&0.414&0.49\\
$N(2100)$&1.885&0.113&0.27\\
$N(2190)$&2.127&0.547&0.29\\ \hline
$\Delta$(1700)&1.762&0.599&0.08\\
$\Delta$(1900)&1.920&0.263&0.38\\
$\Delta$(1905)&1.881&0.327&0.86\\
$\Delta$(1940)&2.057&0.460&0.35\\ 
$\Delta$(2000)&1.752&0.251&0.22\\ \hline
$\phi$(1020)&1.020&0.0045&0.13\\
$h_1$(1170)&1.170&0.36&1\\
$a_1$(1260)&1.230&0.40&0.68\\
$\pi$(1300)&1.300&0.40&0.32\\
$a_2$(1320)&1.318&0.107&0.70\\
$\omega$(1420)&1.419&0.174&1\\
\end{tabular}
\end{table}

\setlength{\topmargin}{-1cm}
\setlength{\textheight}{30cm}
\newpage
\newpage
\newpage
\newpage
\newpage
\newpage

\end{document}